# Characterizing the effects of free carriers in fully-etched, dielectric-clad silicon waveguides

Rajat Sharma, Matthew W. Puckett, Hung-Hsi Lin, Felipe Vallini, and Yeshaiahu Fainman

*Department of Electrical & Computer Engineering, University of California, San Diego, 9500 Gilman Dr, La Jolla, CA 92023*

We theoretically characterize the free-carrier plasma dispersion effect in fully-etched silicon waveguides, with various dielectric material claddings, due to fixed and interface charges at the silicon-dielectric interfaces. The values used for these charges are obtained from the measured capacitance-voltage (C-V) characteristics of $SiO_2$, $SiN_x$, and $Al_2O_3$ thin films deposited on silicon substrates. The effect of the charges on the properties of silicon waveguides is then calculated using the semiconductor physics tool Silvaco in combination with the finite-difference time-domain (FDTD) method solver Lumerical. Our results show that, in addition to being a critical factor in the analysis of such active devices as capacitively-driven silicon modulators, this effect should also be taken into account when considering the propagation losses of passive silicon waveguides.

## I. INTRODUCTION

In the ongoing effort to realize low-loss, high-energy efficiency, and high-bandwidth active device components in nanoscale silicon waveguides, one of the most direct ways to control the propagation of a guided mode is through the application of a capacitive bias electric field. In particular, strained silicon has frequently been shown to exhibit a large second-order nonlinear susceptibility, making it a strong candidate for electro-optic modulation[1-5]. However, recent work suggests that the analysis of semiconductor waveguides subjected to a DC electric field may be more complex than originally anticipated[6,7]. In particular, band-bending and consequent redistribution of carriers at silicon-dielectric interfaces is anticipated to change both the real and the imaginary parts of silicon's refractive index. Additionally, we show that there is a large electric field present in the waveguide even in the absence of an applied bias due to fixed interface charges in the dielectric films. Although these effects are gradually gaining attention within the framework of integrated photonics, more work still needs to be done to decouple free-carrier effects from other distinct properties of silicon waveguides.

In this manuscript, we theoretically quantify the aforementioned changes in lightly doped p-type silicon waveguides, assuming they have been clad with either silicon dioxide, silicon nitride, or aluminum oxide. In addition to possessing distinct indices of refraction, each of the considered cladding materials is known to have different densities of (1) fixed charge and (2) interface trap states at interfaces with (100) silicon[8]. Through our analysis, we find that these differences lead to large changes in the concentration of holes and electrons within the waveguide, even in the absence of a bias electric field. The different changes in carrier concentration are then found to have unique and noteworthy impacts on the modes supported by each of the waveguides under consideration. Furthermore, we show that the changes in carrier concentration in these waveguides in response to externally applied bias electric fields are dominant in determining the value of the field within the waveguide. This last point is particularly important to bear in mind in the characterization of active, capacitively driven silicon modulators.

## II. MEASUREMENT OF DIELECTRIC FILM PROPERTIES

When a semiconductor comes into contact with a dielectric, the semiconductor's valence and conduction band energies may locally deviate from their natural values, leading to perturbations away from the material's bulk carrier concentrations[9]. To determine the extent to which this basic effect occurs, two non-idealities to consider are the concentration of fixed charges within the dielectric and the existence of interface trap states. Whereas the former may shift the semiconductor's response to an externally applied bias voltage one way or the other, the latter may reduce the magnitude of this response altogether. In a silicon waveguide, any changes in the guiding material's carrier concentrations will produce a change in its optical properties, so it is vital to take these effects into consideration.

To determine the fixed charge present at a semiconductor-dielectric interface, the most common technique is to analyze a C-V measurement for a MOS structure consisting of the materials of interest. Because the fixed charge itself has a biasing effect on the carriers within the semiconductor, it tends to offset the response of the MOS structure to external voltages. By observing the experimental flat-band voltage of the device and comparing it to the theoretical value, the concentration of fixed charge in the dielectric may be calculated as[10]:

$$Q_f = C_{dielectric}(\Delta\varphi_{ms} - V_{FB}) \qquad (1)$$

where $C_{dielectric}$ is the dielectric capacitance per unit area, $\Delta\varphi_{ms}$ is the difference in work function between the metal and the semiconductor (and therefore its flat-band voltage in the absence of any dielectric non-idealities), and $V_{FB}$ is the measured flat-band voltage. It is important to note that the total fixed charge within the dielectric may change with the applied voltage due to bulk charge traps within the dielectric, potentially leading to hysteresis in the C-V measurement. But because we are only interested in order-of-magnitude analyses of the previously outlined effects of free carriers, and additionally because we are not considering the application of voltages large enough to induce appreciable hysteresis, we may neglect this complication.

In comparison to fixed charge, interface trap states are not necessarily charged, but rather may become positively or negatively charged in response to any field present at the semiconductor interface. Acceptor-type trap states may take an electron from the semiconductor and thus become negatively charged, and similarly donor-type trap states may give an electron to the semiconductor and become positively charged. The magnitude and sign of interface charge is therefore a function of the voltage applied across the MOS structure, as well as the type of interface state present. Because interface states have long response times, their effects will not affect high-frequency capacitance measurements. As a result, at the bias voltage corresponding to the minimum low-frequency capacitance the interface state density may be calculated as[11]:

$$D_{it} = \frac{C_{dielectric}}{q}\left(\frac{C_{LF}}{C_{dielectric}-C_{LF}} - \frac{C_{HF}}{C_{dielectric}-C_{HF}}\right) \quad (2)$$

where q is the electron charge, $C_{LF}$ is the low-frequency capacitance per unit area of the total MOS structure, $C_{HF}$ is the corresponding high-frequency term.

To obtain these values for silicon dioxide and silicon nitride, we fabricated two MOS structures, one consisting of each material. Using an Oxford Plasma-Enhanced Chemical Vapor Depositor (PECVD), we deposited 150 nm of silicon dioxide and 180 nm of silicon nitride, respectively, on two silicon wafers doped with boron at a concentration of 1e15 cm$^{-3}$. We then fabricated 200 nm-thick aluminum contacts on top of the dielectrics via photolithography and electron beam evaporation. Finally, we briefly annealed both samples at a temperature of 300 °C to improve film quality and carried out low- and high-frequency C-V measurements using an Agilent B1500A Semiconductor Device Analyzer[12]. The C-V curves generated for the two materials are shown in Fig. 1a and 1b. For the case of aluminum oxide, as well as the buried silicon dioxide, the values of interest were taken from the literature[8]. Specific properties of the interface traps such as the electron and hole recombination lifetimes were taken from the literature as well[13,14]. The fixed charge and interface state densities for each of the materials under consideration are listed in Table 1.

TABLE I. Charge densities in dielectric films.

| Material | $Q_f$ (C/cm$^{-2}$) | $D_{it}$ (cm$^{-2}$ eV$^{-1}$) |
|---|---|---|
| PECVD SiO$_2$ | 1(10$^{12}$) | 2.5(10$^{10}$) |
| PECVD SiN$_x$ | 3(10$^{12}$) | 2.1(10$^{10}$) |
| ALD Al$_2$O$_3$ | -2.2(10$^{12}$) | 1.1(10$^{11}$) |
| Buried SiO$_2$ | 4(10$^{10}$) | 1.3(10$^{10}$) |

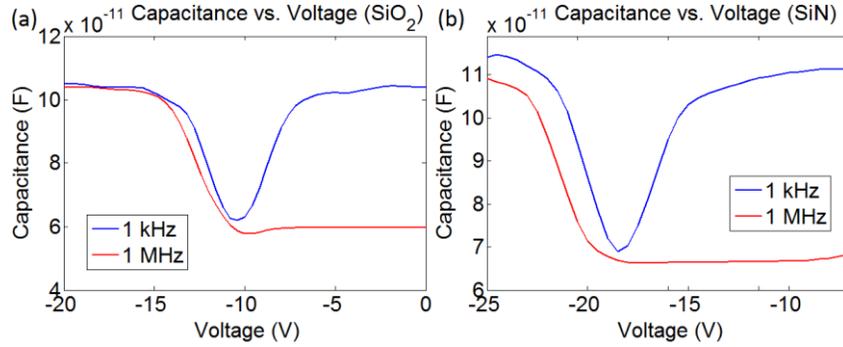

Fig. 1. Measured C-V characteristics showing the low- and high-frequency capacitance as a function of the bias voltage for (a) silicon dioxide and (b) silicon nitride films.

## III. RESULTS AND DISCUSSION

Based on the obtained experimental and theoretical values, we created two-dimensional models using the semiconductor physics tool Silvaco to determine the concentration of holes and electrons across a 250 nm-tall, 500 nm-wide silicon waveguide clad with each of the mentioned dielectrics[15]. By changing the bias voltage applied vertically across the simulation space, we were additionally able to observe how free carriers affected the electric fields induced within the waveguides. Cross-sectional plots of the electron (minority carrier) concentrations are shown in Fig. 2a-2f for both the negatively and positively biased cases, and for each of the materials under consideration. Also, Fig. 2g shows the geometry assumed in the model. For the waveguides with either silicon dioxide or silicon nitride cladding, the positive fixed interface charges push the silicon into depletion, whereas the negative fixed charge inherent to the aluminum oxide, leads to accumulation. Moreover, although the voltages applied across the waveguide lead to changes in carrier concentration, the dominant effect in determining the behavior of the semiconductor is the fixed charge density.

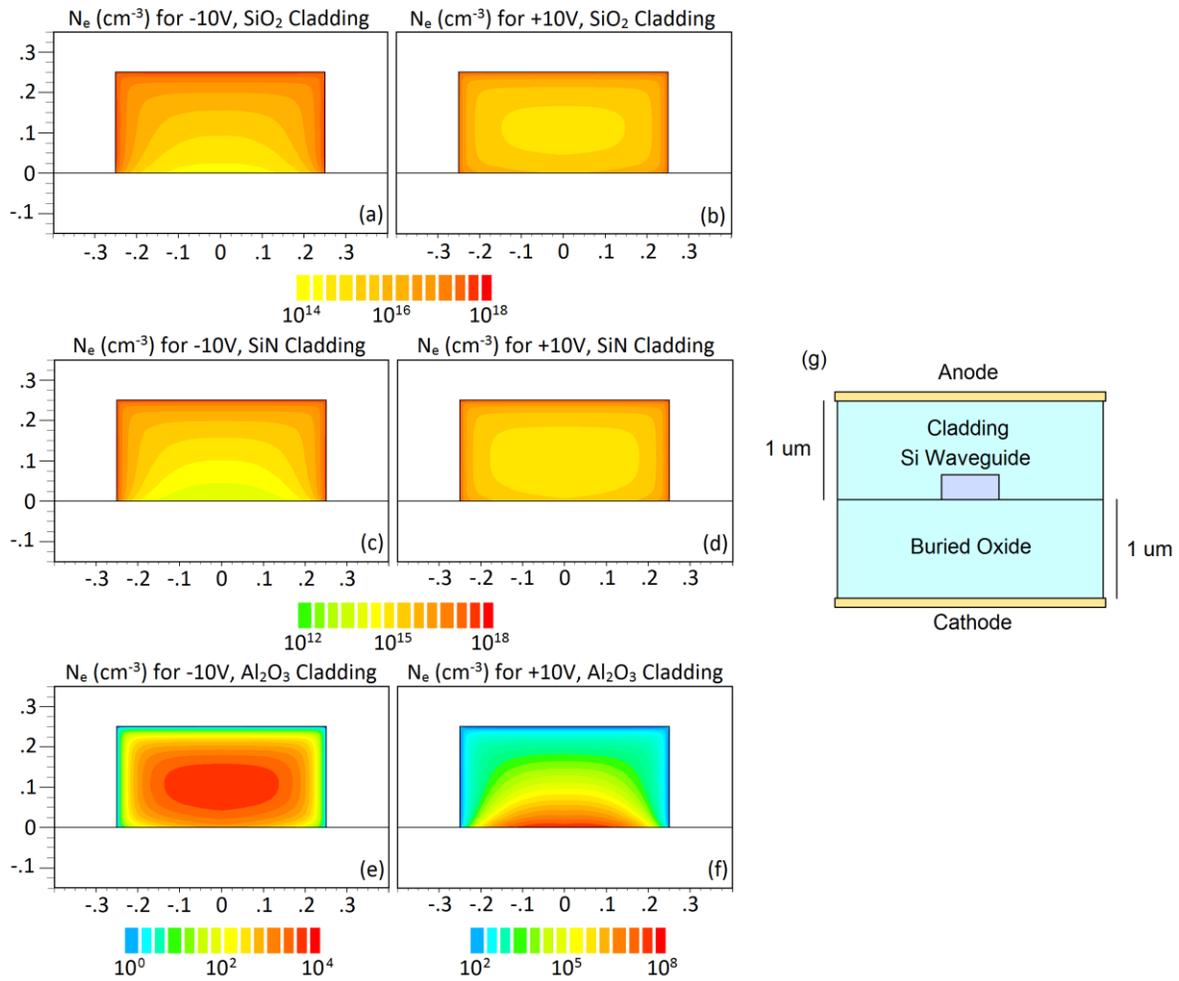

Fig. 2. Electron concentrations across silicon waveguides, assuming vertical bias voltages of either -10 V (left column) or +10 V (right column), when the cladding material is (a,b) silicon dioxide, (c,d) silicon nitride, or (e,f) aluminum oxide. Spatial dimensions are given in units of microns. (g) Illustration of the geometry used in Silvaco.

To further consider the interaction between silicon's free carriers and the electric field in the waveguides, we additionally observed the electrostatic behavior of the waveguide clad with silicon nitride. Our results, shown in Fig. 3, highlight the shielding effect provided by the carriers. When applying a bias of 10 V, the average and the peak value of the electric field (plotted along a y-slice at the center of the waveguide in Fig 3) show minimal changes from their no-bias value, despite the large field values present just beyond the waveguide's boundaries. This illustrates how well free carriers may reduce the interaction of silicon with capacitively applied bias fields.

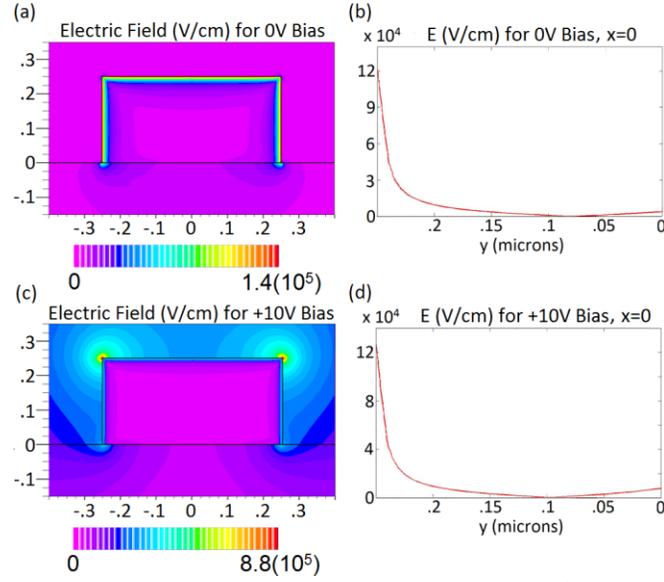

Fig. 3. The electric field distribution within a silicon waveguide, assuming vertical bias voltages of either -10 V (above) or +10 V (below), when the cladding material is silicon nitride. The spatial dimensions are indicated in units of microns.

Following this result, we also investigate the effect of free carriers in silicon on the optical properties of the same set of waveguides. The free-carrier concentrations were translated to spatially dependent deviations away from the unperturbed real and imaginary parts of silicon's index of refraction, given phenomenologically as[16]:

$$\Delta n = -5.4\left(10^{-22} cm^3\right)\Delta N_e^{1.011} - 1.53\left(10^{-18} cm^3\right)\Delta N_h^{.838} \tag{3a}$$

$$\Delta \alpha = 8.88\left(10^{-21} cm^2\right)\Delta N_e^{1.167} + 5.84\left(10^{-20} cm^2\right)\Delta N_h^{1.109} \tag{3b}$$

where $\Delta N_e$ is the perturbation in the concentration of free electrons and $\Delta N_h$ is the corresponding value for holes. We then used the modified values of refractive index with our FDTD solver, the multiphysics software Lumerical, to observe the effect of the carriers on the supported modes of the silicon waveguides[17]. Note that this entire analysis was performed assuming an optical wavelength of 1.55 μm. Our results for both the fundamental TE- and TM-like modes are shown in Fig. 4a-4c. It is evident that the changes in the complex effective index due to plasma dispersion are different for each of the three dielectrics. Aluminum oxide, for example, exhibits deviations in the real part of effective index as large as $2 \times 10^{-5}$ for a 10 V bias, whereas the effect is much smaller in waveguides clad with silicon dioxide. It is also interesting to note that, for each effect, the slope of the curve is opposite in sign for the waveguide clad with aluminum oxide, due to the negative value of the cladding's fixed charge.

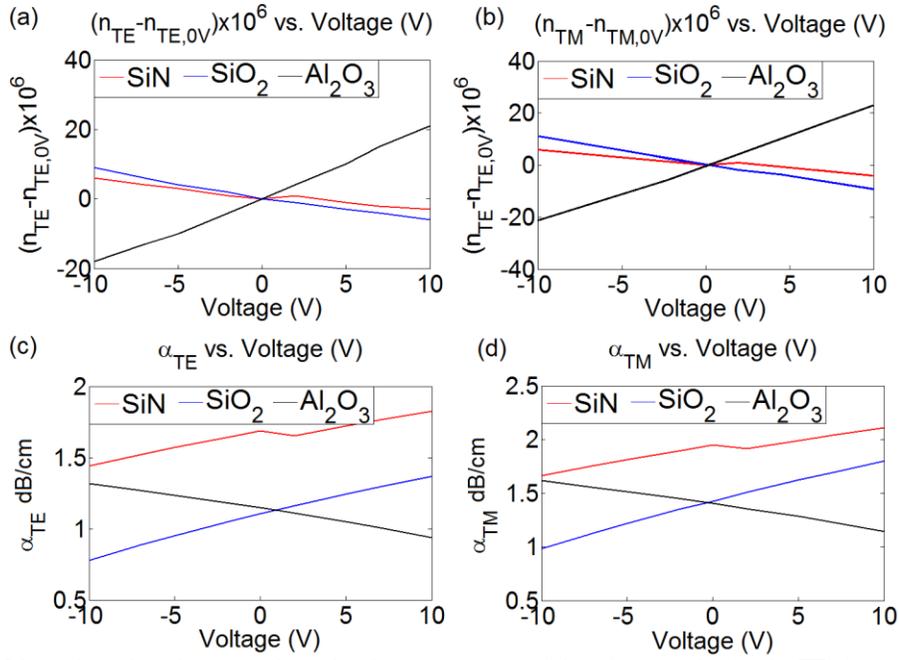

Fig. 4. (Color online) Voltage-dependent changes in the real and imaginary parts of the effective index for the TE- (a,c) and TM-like (b, d) modes, respectively. The legend in the top-right image applies to each of the others as well.

## IV. CONCLUSION

In conclusion, we have shown that silicon waveguides with dielectric claddings exhibit strong spatial deviations in carrier concentration, even in the absence of capacitively applied bias voltages, and that these changes affect the real and imaginary parts of the effective indices for the guided modes. Additionally, we have shown that the same free-carrier effects impact the waveguides' apparent electro-optic effects in response to driving voltages, both because of the changes in local carrier concentration and because of the effects the carriers have on the electric field ultimately induced across the waveguide. In future research efforts, AC measurements may be explored as a tool for decoupling the effects discussed here from other material properties. We suggest that these effects, which we have theoretically quantified, are important to be taken into consideration in the future characterization of both active and passive device components based on silicon waveguides.

## ACKNOWLEDGEMENTS

This work was supported by the Defense Advanced Research Projects Agency (DARPA), the National Science Foundation (NSF), the NSF ERC CIAN, the Office of Naval Research (ONR), the Multidisciplinary University Research Initiative (MURI), and the Cymer Corporation. Sample fabrication and fabrication was performed at the UCSD Nano3 cleanroom facility.